# Modelling techniques for biomolecular networks

**Authors:** Gerhard Mayer[1]

1 Ruhr University Bochum, Faculty of Medicine, Medizinisches Proteom-Center, D-44801 Bochum, Germany

**Abbreviations:**

| | |
|---|---|
| ASP | Answer Set Programming |
| (Open)BEL | Biological Expression Language |
| BioPAX | BIological Pathways eXchange |
| CoD | Coefficient of Determination |
| CoLoMoTo | Consortium for Logical Models and Tools |
| COMBINE | COmputational Modelling in BIology Network |
| EBN | Extended Boolean Network |
| GRN | Genetic Regulatory Network |
| ODE | Ordinary Differential Equation |
| PBN | Probabilistic Boolean Network |
| PDE | Partial Differential Equation |
| PDS | Polynomial Dynamical System / Power Dominating Set |
| RBN | Random Boolean Network / Restricted Boolean Networks |
| SBML | Systems Biology Markup Language |
| STP | Semi-Tensor product |

**Summary:**

First we shortly review the different kinds of network modelling methods for systems biology with an emphasis on the different subtypes of logical models, which we review in more detail. Then we show the advantages of Boolean networks models over more mechanistic modelling types like differential equation techniques. Then follows an overlook about connections between different kinds of models and how they can be converted to each other. We also give a short overview about the mathematical frameworks for modelling of logical networks and list available software packages for logical modelling. Then we give an overview about the available standards and ontologies for storing such logical systems biology models and their results. In the end we give a short review about the difference between quantitative and qualitative models and describe the mathematics that specifically deals with qualitative modelling.

## 1. Introduction

There are plenty modelling techniques from system biology available, which can be applied for modelling the different kinds of biomolecular networks. For newcomers it's not easy to select the appropriate model type. Therefore we review the different network modelling approaches with an emphasis on logical network modelling, what distinguishes this review from others like e.g. [1]. We also introduce the different kinds of mathematical frameworks, which can be used for logical network modelling and computation of system theoretical characteristics and the simulation of network control approaches. These are the semi-tensor product (STP) of matrices [2], Zhegalkin polynomials [3] and the algebraic methods of polynomial dynamical systems (PDS) [3] and model checking [4]. In the end we will mention the various standard formats available for describing, exchanging and archiving systems biological network models data.

## 2. Modelling approaches in systems biology

Systems biology and adequate mathematical modelling of genome-wide resp. proteome-wide biomolecular networks in order to simulate processes in cell biology is still a big challenge and has diverse applications like for instance for network-based biomarker discovery or drug target identification. Molecular interaction networks can be subdivided into metabolic networks, genetic regulatory networks (GRN's) and (small) protein signaling networks [1], whereas the network modelling and analysis of proteome-wide expression data is still in its infancy.

Roughly spoken one can distinguish in a 2x2 scheme discrete from continuous and deterministic from non-deterministic modelling techniques. Each of these 4 subtypes can be further divided according to the update scheme into synchronous vs asynchronous methods. An overview about the main modelling methods in systems biology is given elsewhere in the literature [1,5-10] and summarized in Table 1. It should be noted that there are also plenty methods described in the literature, which combine two or even more of these pure modelling methods into hybrid modelling schemes. One example are the Stochastic Discrete Dynamical Systems (SDDS) [11] combining the Polynomial Dynamical Systems (PDS) with the stochastic Gillespie models or the ANIMO (Analysis of Networks with Interactive Modeling) method, which lies between ODE's and fuzzy logic models [12].

Table 1: Overview about modelling approaches in systems biology. The logical models are reviewed in more detail in section 2.1 and Table 2.

| Modelling approach | Short description | References |
|---|---|---|
| Logical models (Boolean networks) | The (virtual) interactions between the proteins leading to the observed co-expression are expressed as Boolean functions [13]. See Table 2 for a detailed classification of logical network modelling approaches. | [14,15] |
| Petri nets (PN) | A special kind of bipartite graph with two kinds of nodes: places and transitions. If a transition fires a token is moved from the input places to the output places; asynchronous and non-deterministic. | [16,17] |
| Polynomial dynamical systems (PDS) | An algebraic based method which is a special kind of a sequential FDS (Finite Dynamical System) over a finite field. Each transition function is an element of a polynomial ring over the finite field. Uses enhanced fast methods from computer algebra and computational algebraic geometry (rooting in Buchberger algorithm) to calculate the Gröbner bases of ideals in such rings [18] and an ideal is a set of polynomials, which is closed under polynomial | [20] |

| | combination [19]. | |
|---|---|---|
| Differential equation models (ODE and PDE) | ODEs (Ordinal Differential Equations) typically are used to model the (time) dynamics behavior of networks, whereas PDEs (Partial Differential Equations) are used to model the behavior in space and time, so that pattern formation can be modelled. Such spatiotemporal Diffusion-Reaction Systems exhibit self-organizing pattern formation, generally described by the general local activity principle [21], which explains the cause of complexity and self-organization in nature. | [22,23] |
| (Dynamic) Bayesian models | Probabilistic method allowing to incorporation of prior information by using Bayes Theorem. Problematic can be the derivation of the direction of an interaction. | [24,25] |
| Process calculi | Models the network interactions as a sequence of stochastically occurring communication events instead of states. Different kinds of process calculi are e.g. the pi-calculus and Beta-binders. Scalability to larger models is bad. Process calculi can be seen as a subgroup of the rule-based approaches. | [26] |
| (Dynamic) Cellular automata | A grid of cells with a finite number of states which change according to rules in discrete steps. The rules determine the new state of a cell in dependence of the cells current state and the state of the neighboring cells. Mainly used for simulation of aspects of spatial dynamics, e.g. diffusion processes and pattern formation [27]. | [28] |
| Interacting state machines | Qualitative high-level hierarchical modelling of objects (e.g. cells) by its parts (e.g. genes and proteins) at different levels. Therefore allows compositional multiscale models with synchronous or asynchronous state changes and visualization by state charts. One advantage of these models is that they allow the application of model checking techniques [29]. | [5] |
| Finite State Linear Model (FSML) | Combines continuous (e.g. protein concentration) with discrete (e.g. promoter regions with a finite number of states) modelling aspects. | [30] |

| Agent-based models (ABM) | Originally developed in the social sciences and economics. Simulates the micro-behavior and the interactions of autonomous agents (e.g. genes, mRNAs (siRNA, miRNA, lncRNA), proteins, transcription factors) and studies their effects on the macro level, i.e. the whole system (e.g. the cell). | [10,31] |
|---|---|---|
| Rule – based models | Here the molecular interactions are modelled with local rules, which can be applied even if there is no explicit network structure available, i.e. the network inference step is not necessary, so that these network-free approaches bypass the combinatorial burden of network inference. | [32] |
| Piecewise-linear differential equation models (PLDE) | Model consists of a piecewise-linear approximation of differential equations by means of step functions together with a set of inequality constraints for the parameter values. | [33] |
| Stochastic models | Models based on the Gillespie algorithm for solving the chemical master equation, which gives the probability that a molecular species will have a specified molecular population resp. concentration at a given future time [34]. The Gillespie method is the computationally most expensive approach. If one has a low number of molecules or one wants to model molecular crowding effects, then the stochastic approach is the method of choice [35]. | [36,37] |
| State Space Model (SSM) | Linear [38] or non-linear [39] modelling approach using an abstract space of states in combination with a diverse set of algorithms, among them are Bayesian or other statistical based methods, autoregressive models and Kalman filtering. | [38,39] |
| MP-Systems | Metabolic P-systems, for modeling the dynamics of metabolism and signal transduction based on the membrane computing model, which was shown to be capable of solving NP-complete problems in polynomial time [40]. | [41] |

Each of the listed models has its advantages and disadvantages. The decision for a model type depends on several factors, e.g. the size of the model, the availability of data for the identification of the model or on the intended level of insight (qualitative or quantitative) one wants to gain, as well as on the maturity of the underlying mathematical framework. For Boolean networks, for polynomial dynamical systems, for differential equation models and for piecewise-linear differential equation models a rigid mathematical framework exists. Whereas for Petri nets, cellular automata, interacting state machines, agent-based and rule-based methods some arbitrary degrees of freedom are introduced by the rules, how to model exact the interaction between the molecular species resp. when to fire a transition. If one wants to simulate spatial aspects, then one should use PDE or cellular automata models are the right choice, whereas for multiscale modelling interacting state machines are a suitable choice. Process calculi are only valuable for application to reasonable small models. Other models need the availability of some special data, e.g. some prior knowledge is required for Bayesian models and kinetic data are required to apply stochastic models based on the Gillespie algorithm.

The use of ODE/PDE for complex models has serious drawbacks rooting in non-identifiable parameter sets [42] because of incomplete and noisy data. There are often experimental constraints, which do not allow measuring all the needed parameter values [43] necessary for calibration of the ODE/PDE models. As a consequence, at least some of these parameters must be estimated, which often leads to model overfitting. Also very detailed ODE/PDE models often comprise differential equations, where no analytical solution is available. As a consequence application of numerical solution methods is indispensable [44], which frequently leads to oscillating model behavior and/or to long calculation runtimes. Often one has to resort to integer arithmetic [45], which slows down the algorithm performance, since rounding errors in floating point arithmetic leads to wrong numerical values. Another drawback is that differential equation based models are sloppy, i.e. they can be tuned to reality by adjusting one key parameter per stiff direction, independently of how reliably the other parameters are estimated [46]. In addition, the calibration of such models requires time series data, which usually are not measured in Proteomics studies. The biggest drawback is that quantitative data in Proteomics are mostly given in relative terms and not in absolute quantities like amount or concentration. Therefore it seems that the theoretically attainable precision of a model should be adapted to the accurateness of the noisy measurements. Such logical based models [47] allow one to simulate reliably at least the qualitative network dynamics. In [48] it was shown that Boolean models can be seen as coarse-grained limit case of differential equation models and that they are able to reproduce the results of them despite that simplification. In summary, it is not feasible to apply such exhaustive differential equation based techniques for modeling of more complex biological systems like proteome-wide data sets on the molecular level. The reason is that the number of parameters for differential equation models is larger than for Boolean models, since they need additional parameters for the reaction kinetics [49]. In addition the Boolean models are much easier to interpret, since they are more intuitively understandable [50].

## 2.1 Logical network modelling approaches

Because of the advantages of logical models in comparison to differential equation models [47], we will in the following give a detailed classification of the different kinds of logical modelling approaches (Table 2).

Table 2: Overview about systems biological logical network modeling approaches.

| Logical modelling approach | Short description | References |
|---|---|---|
| Random Boolean Networks (RBN) | The most general type of Boolean Network is an exemplar drawn from the set of all possible Boolean networks without any restrictions, whereas the networks occurring in biology, called simply Boolean Networks, differing from RBNs by preferring network structures with a small amount of attractors and large basins of attraction [51]. Gershenson [52] gives a detailed classification of random Boolean networks according to the update schemes (deterministic, random, synchronous or asynchronous). The non-deterministic asynchronous RBNs are also called contextual RBNs [53].RBNs differ from naturally occurring Boolean networks also by some topologically aspects [54], even if the scale-freeness and power law distribution of the node connectivity cannot always be confirmed [55]. | [50] |
| Boolean networks (BN) | The nodes correspond to proteins and the edges to (virtual) regulatory relationship between them. Networks with N nodes and an in-degree of K (K<=N) for every node are called N-K-nets. For modelling of biological networks often K is restricted to a small number (2<=K<=10) [56], whereas for RBNs K can be equal to N. Special subtypes of BNs are the temporal BNs [57], which take into account regulatory delays and master-slave Boolean | [61,62] |
| | networks [58,59]. A variant of temporal BNs are time-delay and diffusion-reaction BNs [60].<br><br>One big advantage is that the Boolean networks are easy to interpret. | |
| Extended Boolean Networks (EBN) | Extended Boolean networks are one possibility to integrate known post-transcriptional regulation processes exerted for instance by miRNA. Here the set of nodes is extended to include more than one type of biomolecules, namely beside the gene nodes there are also nodes for mRNA-protein pairs and nodes for miRNA. In addition there are 4 possible edge types: gene–gene, protein–gene, gene–miRNA and miRNA-protein. | [63] |
| Biological Function Network (BFN) | Two-step reverse engineering method based on a Hidden Markov Model (HMM) with a worst-case time-complexity of $O(n^3)$. | [64] |
| Fuzzy and Multi-Valued Boolean Networks, Generalized Logical Networks (GLN), Quantitative Logic Models (QLM), constrained fuzzy logic models (cFL) | Finer-grained version of a Boolean network. Overcomes the Boolean limitation that a protein is either expressed or not expressed. In the fuzzy version every continuous value from between the interval 0 to 1 is allowed for the influence strength of a protein to the expression of the other. In the multi-valued version k equally distributed discrete expression values in the range from 0 to 1 are allowed. A variant of these multi-valued networks are the GLNs [65].<br><br>QLM and cFL are prior knowledge networks described by graded values of protein activation. | [66,67] |
| AND-NOT networks | AND-NOT (or conjunctive) Boolean networks are simplifications. It was shown that every finite dynamical system and thereby Boolean network can be rewritten as AND-NOT network with similar dynamic properties and the same number of steady states by introducing some additional nodes. Note that AND-NOT functions are a particular case of nested canalizing functions [68]. The advantage is that AND-NOT networks can be handled with up to 1 million nodes. | [69,70] |

| Probabilistic Boolean Networks (PBN) | Probabilistic Boolean network introduce a stochastic element. They can be seen as a set of Boolean networks models, where a probability distribution governs the switching between them. The rationale behind is that one tries to model the uncertainty resulting from the network inference process, because one typically has only a limited number of samples (examples) relative to the number of genes [71]. In practice one infers a number of good simple predictors from the experimental data for each target gene and probabilistically synthesize a real predictor such that each predictor's contribution is proportional to its determinative potential, as measured by the Coefficient of Determination (CoD) [71]. Namelythe state transitions between the genes are represented by a list of Boolean functions instead of only one Boolean function. The master-slave PBNs [59] are a subtype of the PBNs. | [72] |
|---|---|---|
| Stochastic Boolean Networks (SBN) | SBNs are an improved version of PBN's, which allow to more efficiently compute the state transition matrix in $O(nL2^n)$ compared with $O(nN2^{2n})$ in the general resp. $O(nN2^n)$ in the sparse matrix case for PBN's, where n is the number of genes, N the number of Boolean networks and L a factor determined by the stochastic sequence length. L increases polynomial with n and therefore is typically smaller than N, which increases exponentially with then number of genes n. | [73] |
| Restricted Boolean Networks (RBN) | As explained for the PBN method, there are due to noise and the small amount of samples several networks, which can explain a given data set. Analyzing the similarities between these networks one can derive confidence measures for the relationship between nodes. RBN's are restricted Boolean networks derived from time series data, where by constraints only a subset of all possible Boolean functions between the nodes is considered, based also on assumptions derived from pairs, double pairs or triples of states in the | [74] |

| | time series data set. For instance only an active gene at time t can regulate other genes at time t+1. | |
| --- | --- | --- |
| Threshold Boolean networks (TBN) | In TBN's the expression value of a protein is computed by the sum of influences from all other proteins in the network. If this sum is above a threshold, the target protein is activated and if the sum is below the threshold, the target protein is deactivated; otherwise the expression of the target protein remains in its current state. The advantage is that compared with traditional BN's specified by the truth tables of the Boolean functions, much fewer parameters are required for TBN's: The truth table for a N-K net has $\sum_{i=1}^{N} 2^K$ parameters, whereas a TBN requires only $\sum_{i=1}^{N} K$ parameters. This makes them excellent candidates for exploratory studies, where one don't want to include prior information, required to make the inference of large networks computational feasible. | [75] |
| Markov logic networks (MLN) | This type of models is a probabilistic graphical model in combination with first-order logic. | [76] |
| Zhegalkin Polynomials, Reed Muller Forms | Zhegalkin polynomials, also called Reed-Muller Forms are algebraic normal forms of Boolean functions and allow a continuous representation them. The computation of these normal forms is based on list-decoding [77] or tensor decomposition techniques [78]. Recently also a Gröbner-free method for computing these normal forms was developed [79]. | [80] |

RBNs are mainly used for studying the physical properties of general network and are therefore not relevant for biological applications. The other types logical models can all be used to model protein signaling pathways, which are rather small networks. For bigger proteome-wide networks the network inference process can be computationally quite expensive, especially, when the in-degree K is bigger. For larger networks one should guide the inference process by prior knowledge, from literature or from protein interaction of transcription factor databases, which can be done e.g. by regularization approaches [81] or by introducing constraints to restrict the search space. Other possibilities are the use of AND-NOT networks, or threshold Boolean networks, which work very well even for larger networks or one confines the inference to networks with a low in-degree K. This is possible since it was shown that for the yeast proteome net an in-degree of K=5 is sufficient [82], since 93% of the genes were regulated by only 1-4 proteins [83]. Such conjunctive Boolean networks make sense, since the AND functions correspond to the synergistic regulation of a molecule by several factors [84].

### 3. Dependencies between the different modelling methods

There are many connections and interdependencies between the different modelling approaches, so that there are several methods for conversion between different model types [1]. An overview about such model conversions and equivalencies gives table 3.

Table 3: Overview about model type conversion approaches resp. model type equivalencies.

| (Source) model type | (Target) model type | Reference |
|---|---|---|
| Ordinary Differential Equations | Boolean networks | [48] |
| Boolean networks | Petri nets | [16,85] |
| Boolean networks | Constraint-based models | [86] |
| Boolean networks | Ordinary Differential Equations | [87] |
| Constraint-based models | Ordinary Differential equations | [88] |
| Petri nets | Ordinary Differential Equations | [89] |
| Petri nets | State-transition automaton | [90] |
| Process algebra | Ordinary Differential Equations | [91] |
| Rule-based models | Ordinary Differential Equations | [92] |
| Process calculus | Colored Petri nets, Ordinary Differential Equations, continuous time Markov chains | [93] |
| Discrete model | Piecewise linear model | [94] |
| Probabilistic Boolean network | Dynamic Bayesian networks | [95] |
| Boolean Networks | Polynomial Dynamical Systems | [96] |

| Probabilistic Boolean Networks | Polynomial Dynamical Systems | [97] |
|---|---|---|
| Petri nets | Polynomial Dynamical Systems | [98] |
| MP systems | Hybrid Functional Petri Nets | [99] |

Such conversions can alleviate the human interpretability of results or can be used for modelling some orthogonal concepts, which cannot be modelled adequately by one model type. For instance the interpretability of Boolean functions in their polynomial form can be difficult to interpret and Petri nets can be used for asynchronous modelling, which is not possible in polynomial dynamical systems. The PDS are namely subtypes of Sequential Dynamical Systems (SDS's) [96], which in turn are a class of Graph Dynamical Systems (GDS), which are the most general framework for discrete logical graph models. They consist of a finite graph, where each node can be in a finite set of states. These states are updated by update functions following an update scheme, which specifies the selection of the next update function. When one expresses the update functions by a polynomial function over a finite field one speaks of polynomial dynamical systems (PDS) [97]. Therefore the algebraic method of polynomial dynamical systems can also be seen as a logical modelling method, since a Boolean network can easily be converted into a PDS by representing the Boolean functions as square-free polynomials with coefficients in the finite field $\square/2 = $ [96], where AND is replaced by multiplication, OR by addition and NOT by the addition of 1, since the arithmetic is over the Boolean field:

$$\text{AND}(X, Y) \quad = X \wedge Y \quad := XY$$

$$\text{OR}(X, Y) \quad = X \vee Y \quad := X + Y + XY$$

$$\text{NOT}(X) \quad\quad\quad := X + 1$$

These PDS's allow not only the modelling of Boolean Networks and Probabilistic Boolean Networks [97], but are a very general framework under which most of the discrete modelling techniques like e.g. finite state machines, interacting state machines, dynamic Bayesian networks, agent-based approaches and Petri nets [98] can be subsumed. It's also clear that the PDS's have close relationship to the Zhegalkin polynomial based method and the methods based on the semi-tensor product (STP) of matrices [2]. Whereas in the normal matrix product, describing linear transformations,, the elements of the result matrix are calculated as a sum of element-to-element products, and the elements of a tensor product result tensor are the product of one element with a whole matrix, for the semi-tensor product the elements of the resulting matrix are calculated by the product of one element times a block of the other matrix [100].

## 4. Mathematical frameworks for modelling of Boolean networks

There exits mainly four mathematical frameworks, all of them based on a sound mathematical basis, for modelling of Boolean networks and for calculating their system theoretic descriptors. The most general framework for logical models is the PDS (see section 2). Two other closely related frameworks are the STP and Zhegalkin polynomials. Which framework performs best for a given task is still an open question, even if the STP method is currently the most used method for solving Boolean network control problems. Therefore we shortly introduce here the application of all four frameworks to problems of logical network modelling.

### 4.1 Computation with STP

The STP is a generalization of the conventional matrix product requiring no dimension matching condition [2,101]. In [102] it was shown that it can be used for the modelling of Boolean networks. A big advantage of the STP approach is that it can also be used for multi-valued, mix-valued and fuzzy versions of the Boolean network model [103]. Also asynchronous networks can be modelled with the STP method [104]. The STP approach allows the analytical computation of typical system-theoretic descriptors as the number of cycles and fixed points, the basin of each attractor and the transient period for all points to enter these attractors [102,105]. The identification of a Boolean network from experimental data via the STP method is possible [106] and algorithms for the optimal control using the STP were developed [107]. A transformation of Boolean control networks into the Kalman decomposition - a standard form making clear the observables and controllables of the network – is given in [108] and also the control of multi-value logical networks is possible with the STP method [109].

### 4.2 Computation with PDS

The PDS framework is based on polynomial algebra and is the most general framework for the logical modeling of networks. Fast efficient algorithms for solving such polynomial systems, based on Gröbner basis computation [18,110] as well as a Gröbner-free method [79] were developed. There are network inference procedures using the PDS approach available [97,103], and also PDS-based methods for the steady state analysis and derivation of system-theoretic characteristics of Boolean networks [70,84,111] were developed.

### 4.3. Zhegalkin polynomials

The use of Zhegalkin polynomials (Reed-Muller forms) was originally developed in electrical engineering and adapted in [112] to gene expression modelling. These Zhegalkin polynomials are a polynomial on a finite field and are therefore a subclass of the PDSs. They are also related to the STP approach and as shown in [78] the Zhegalkin polynomials are equivalent to Boolean Kruskal tensors and to orthogonal ternary vector lists (OTVL), an approach which is very efficient and widespread used in digital logic design. The inference of the Zhegalkin polynomials using a mixed integer quadratic program (MIQP) allows the incorporation of prior knowledge [80]. An efficient algorithm for the polynomial interpolation using a Boolean tree-based data structure is described in [113].

### 4.4 Model checking

Model checking [116] is a method originally developed for verifying the correctness of electronic circuit designs, which can also be used to model biomolecular networks [4]. The technique does an exhaustive exploration of the networks state-space and verifies that it always adheres to a set of requirements and the model consistency is checked with respect to the experimental data [114]. The model implementations make use of BDD's as efficient data structure [115]. There are model checking algorithms available for computing steady states [116], for finding attractors [117] and for the control of Boolean networks [118] and PBNs [119].

### 5. Available software tools for logical modelling, standards and ontologies for logical systems biology models and simulations

Some software packages available for logical modelling and analysis of systems biology networks are summarized in the supplementary table 1. Many of them are also linked on the ColoMoTo web site http://www.colomoto.org/software/index.html. A list of proprietary formats for storing Boolean models can be found at http://colomoto.org/biolqm/doc/formats.html.

Standards are very important for data exchange and the interoperability of models as well as for the documentation and reproducibility of simulation results. Thereby standards are helpful in fulfilling the FAIR guiding principles [120] for data management. Table 5 lists minimum information guidelines, standards, ontologies and converters relevant for logical systems biology models. A comprehensive overview about systems biology and corresponding visualization standards is given in detail in [121]. The common basis for these standards are the minimum information guidelines documented by MIRIAM (Minimum Information Requested In the Annotation of Models) [122], MIASE (Minimum Information About Simulation Experiments) [123] describing the minimal information required for describing a systems biology model resp. simulation. MIMIP (Minimal Information for Model inference and Parametrization) is a further guideline, currently under development by the COmputational Modelling in BIology NEtwork (COMBINE) [124] (http://co.mbine.org), an initiative defining and promoting computational modelling standards and protocols for systems biology. They defined amongst others the Identifiers.org and MIRIAM Registry [125], allowing the permanent, unique and unambiguous access to models stored on the web via the use of URIs (Uniform Resource Identifiers). Their central format is the COMBINE archive format OMEX (Open Modeling EXchange) [126], which allows the storage of all relevant standardized data files belonging to one model, together with a describing metadata file in RDF (Resource Description Framework) format in one compressed .zip file. This .zip file contains a XML-based manifest file describing the content of the whole .zip file.

CoLoMoTo (Consortium for Logical Models and Tools, http://www.colomoto.org) [127] is an initiative which amongst others is active in defining standards and providing tools relevant especially for logical modelling approaches. They defined the SBML qual format [128], allowing the representation of multivalued logical models.

Terms from ontologies are used to semantically annotate biological and biomedical [129] and proteomics data [130,131]. Important ontologies for systems biology are for example KiSAO (Kinetic Simulation Algorithm Ontology) for describing the simulation algorithms and their parameters, TEDDY (TErminology for the Description of DYnamics), which allows the annotation of the output of simulation runs and the dynamical system behavior. SBO (Systems Biology Ontology) [132] is suited for describing the entities, their role and the model parameters together with metadata describing a systems biology model. MAMO (MAthematical Modelling Ontology) allows specifying the type and characteristics of the used modelling framework (e.g. logical or continuous) and variables. A controlled vocabulary especially for describing logical models is currently under development by the CoLoMoTo consortium [127].

Table 5: Minimum information guidelines, standards, ontologies and converters relevant for logical systems biology models and simulations

| Standard / Ontology | Short description / URL (Last accessed 29th February 2020) | Reference |
|---|---|---|
| **Minimum information guidelines:** | | |
| MIDAS | Minimum Information for Data Analysis in Systems Biology; also an Excel-based format | [133] |
| MIASE | Minimum Information About a Simulation Experiment, guidelines for reproducibly documenting the results of simulation experiments.<br><br>http://biomodels.net/miase/ | [123] |
| MIMIP | Minimal Information for Model Identification and Parametrization. | [124] |
| MIRIAM | Minimum Information Required In the Annotation of Models, a set of guidelines for the consistent annotation of systems biology models.<br><br>http://co.mbine.org/standards/miriam | [122] |
| **Standard formats:** | | |
| SBML | Systems Biology Markup Language, a XML-based interchange format for quantitative computer models of biological processes.<br><br>http://sbml.org/Main_Page | [134] |
| SBML qual | SBML extension package for qualitative network models based on regulatory or influence graphs; well suited for the representation of logical models.<br><br>http://sbml.org/Documents/Specifications/SBML_Level_3/Packages/qual | [128] |
| CellML | Cell Markup Language, a XML-based language for storing and exchange of computer-based quantitative mathematical cell models enabling easy reuse of model components; better suited for continuous models including kinetic details and for higher-level models.<br><br>https://www.cellml.org | [135] |
| BCML | Biological Connection Markup Language, a data format to represent biological pathways in consideration of the organism, tissue and cell type as well as on the physiological, pathological and experimental conditions.<br><br>http://www.compbiotoolbox.fmach.it/BCMLdocs/index.html | [136] |

| BDML | Biological Dynamics Markup Language, an open format for representing quantitative data of biological dynamics.<br><br>http://ssbd.qbic.riken.jp/bdml/ | [137] |
|---|---|---|
| BiSDL | Biology System Description Language, Nets-within-Nets formalism (NWN) | [138] |
| SED-ML | Simulation Experiment Description Markup Language is an encoding of simulation setups in order to ensure exchangeability and reproducibility of time course and steady state simulation experiments.<br><br>http://sed-ml.org | [139] |
| SBRML | Systems Biology Results Markup Language, a markup language for storing the quantitative results of simulations runs of SBML models.<br><br>http://www.comp-sys-bio.org/SBRML.html | [140] |
| SBtab | SBtab is a set of conventions about data tables and spreadsheet files, for use in Systems Biology data exchange.<br><br>http://www.sbtab.net/index.html | --- |
| OMEX | Open Modeling Exchange, a COMBINE archive format containing all files belonging to a model together with a manifest files describing the content in a single .zip file.<br><br>http://co.mbine.org/standards/omex | [126] |
| OpenBEL | Open Biological Expression Language. Uses a triple-based modelling language (subject-predicate-object); suitable especially for causal network models. Fits well for logical models, since BEL allows representing qualitative causal and correlative relationships [141].<br><br>http://www.openbel.org | [142] |
| **Graphical standards:** | | |
| GPML | Graphical Pathway Markup Language<br><br>http://www.pathvisio.org/gpml/ | [143] |
| BioPAX | BIOlogical PAthways eXchange, a language for the integration, exchange, | [144] |
| | visualization and analysis of biological pathway data defined in OWL (Web Ontology Language) / DL [164]. The interactions can be biochemical reactions, binding interactions or regulatory relationships. BioPAX is subdivided into several levels:<br><br>• Level 1:    metabolic pathways<br><br>• Level 2:    molecular interaction networks<br><br>• Level 3:    signal transduction network and gene expression regulation<br><br>• Level 4:    genetic networks / pathways<br><br>• Level 5+:  networks of abstract relationships, e.g. cell–level interactions, environment<br><br>BioPAX is suited especially for data exchange between software and databases.<br><br>http://www.biopax.org | |
| SBGN | Systems Biology Graphical Notation, a graphical notation for biological process maps. It consists of three different languages for different use cases:<br><br>• Process Descriptions (PD): a bipartite graph for the detailed description of biological processes like e.g. metabolic pathways including the kinetics.<br><br>• Entity Relationship (ER) diagrams for describing the biological entities linked by relations, i.e. fits best for rule-based models.<br><br>• Activity flowcharts (AF) describing the activity flow in a biological system. Fits well for logical models, since the detailed mechanistic knowledge is not required.<br><br>For implementing a SBGN diagram one can use either the SBGN-ML markup language [145] or one can resort to SBML layout extension [146].<br><br>http://www.sbgn.org/Main_Page<br><br>http://www.sbgn.org/LibSBGN/Exchange_Format<br><br>http://otto.bioquant.uni-heidelberg.de/sbml/ | [147] |
| **Ontologies:** | | |
| EFO | Experimental Factor Ontology: describes experimental variables like e.g. disease, cell line, organism parts, developmental stage, species, chemical compounds … <br> http://www.ebi.ac.uk/efo/ | [148] |
| KiSAO | Kinetic Simulation Algorithm Ontology for the description of existing algorithms and their interrelationships through their characteristics and parameters. <br> http://co.mbine.org/standards/kisao | [132] |
| MAMO | MAthematical Modelling Ontology is an ontology describing and classifying mathematical models used in the life sciences. <br> http://sourceforge.net/p/mamo-ontology/wiki/Home/ | --- |
| SBO | Systems Biology Ontology, a set of controlled vocabularies of terms for use in Systems Biology. <br> http://www.ebi.ac.uk/sbo/main/ | [132] |
| SBPAX | Systems Biology Pathway Exchange, an extension to BioPAX to include systems biology information into BioPAX models. <br> http://www.biopax.org/mediawiki/index.php/SBPAX3 | [149] |
| TEDDY | TErminology for the Description of Dynamics is an ontology describing the dynamical behaviors, observable dynamical phenomena, and control elements of bio-models in Systems Biology and Synthetic Biology. <br> http://co.mbine.org/specifications/teddy | [132] |
| **Converters:** | http://sbml.org/Software/Converters | |
| SBFC | Systems Biology Format Converter, a generic framework for Systems Biology model format conversions. <br> https://www.ebi.ac.uk/biomodels/tools/converters/ <br> http://sbfc.sourceforge.net/mediawiki/index.php/Main_Page | --- |

A big advantage of such standards is that one can develop dedicated software packages that can run simulations of models stored in a proper standard format, e.g. the systems biology core algorithm [150] can execute models in SBML format. Also programming libraries for accessing and writing SBML files [151] and for simulation of SBML models [152] are available. What is missing until now is an engine for simulations of logical models stored in SBML qual or OpenBEL. As shown, PDSs and STP are ideal candidates for such an execution framework for logical models.

What is missing until now is a quality standard for documenting network validation results, what is urgently required for translating network-based medicine towards the clinics. Applying logical modelling in the clinical practice would in addition require the adherence to regulations like GCLP (Good Clinical Laboratory Practice) [153], and raises also a lot of other questions similar to the ones reviewed in [154] for genomic medicine. It would also require the authentication of all generated files by certificates like for instance X.509. A specialized model repository of all relevant reference networks in human would also be very important.

For the input data for biological network inference one can either use proprietary files like e.g. from MaxQuant [155], or standard data formats for protein quantification data like mzTab-P [156], which is already supported by the PRIDE repository [157]. Another quantitative format is mzQuantML [158], for which it's also expected that it is in future supported by ProteomeXchange [159] and PRIDE for complete quantitative submissions.

The use of the logical modelling techniques for applications in systems biology and systems medicine are discussed in detail in [56] together with the proper inference methods, the validation of the inferred networks, the derivation and meaning of system theoretic descriptors and the application of control approaches for network-based biomarkers, drug target discovery and personalized therapy planning simulations.

## 6. Network inference, validation and control

The inference of the Boolean networks is still very challenging due to the high computational complexity and enormous amount of experimental data for finding an exact solution [56] based on expression values alone. Therefore for the inference of Boolean networks one can either restrict the type of the Boolean network, e.g. to AND-NOT (conjunctive) or disjunctive networks [69], use PAC (Probably Approximately Correct) machine learning methods [160] for the inference. By this restriction of the allowed Boolean functions to conjunctive resp. disjunctive normal forms, which can be learned by PAC learning algorithms, the data requirements are bounded by a polynomial example set size [161].

Or one uses heuristics to simplify and/or prior information to guide the inference process. For instance one can restrict the complete set of multivariate comparisons by setting a threshold for the in-degree of the network nodes [56]. Then the inference procedure uses the expression values to determine the similarity between the expression profiles of two stable network states based e.g. on correlation or information theoretic measures like the CoD or mutual information [56]. Other methods use nonlinear [162] or integer linear programming [163] with constraints, which are either derived from experimental data, from literature [164], from sparsity assumptions [165] or from regulatory constraints [166]. A typical sparse network assumption would be to take into account only the k strongest of the n possible input connections, where k << n and n is the number of network nodes, i.e. to derive functions $f:[0,1]^n->[0,1]$, which are called a *k-juntas*, if they depend on at most k of the input coordinates. Another method is to include prior knowledge for restricting the state space, e.g. using information about molecular interactions and/or pathways, from which one can derive constraints regarding the direct protein connectivity stored in general pathway or molecular interaction databases as reviewed in [167-170]. Or one can utilize information from specialized disease databases like e.g. NeuroDNet for neurodegenerative networks [171] or CancerNet for cancer [172]. Other useful prior knowledge are compartmental information or restraints on the network edge types as in extended Boolean networks [63].

After the inference, the network should be checked for validity, e.g. by iterative pruning to predict missing expression values [173] caused by experimentally noisy and incomplete data. Another pruning approach is PRUNET, which iteratively optimizes the match between the predicted attractors of the inferred network with Boolean representations of two stable phenotypic steady states [174]. A further method for assessing the biological validity uses an interaction relevance distance measure matching the inferred network against the a priori given pathway information, which is even capable of taking into account weak links [175].

Control methods in general are excellently reviewed by *Liu and Barabási* [176] and an overview about control strategies for Boolean networks is already given in the review of Mayer *et.al.* [56]. One goal of network modelling and control is to simulate the effect of external control perturbations, e.g. therapeutic measures on the network behavior, especially on the switching to other phenotypic states, which correspond to attractors and which can be visualized by Derrida plots [56]. Schwab and Kestler described a tool, by one can iteratively screen for perturbations, which change the attractor structure of the network [177] in a desired way.

## 7. Open problems and outlook

Boolean models are intuitively easy to understand, but many computational aspects in their reconstruction, analysis and control are NP-complete [178]. Therefore even the use of reconfigurable or special hardware like FPGA's (Field

Programmable Gate Arrays) [179], ASIC's (Application Specific Integrated Circuits) or GPU's (Graphical Processing Units) does not help for an exact simulation of Boolean networks. Therefore the use of approximations, sparsity assumptions and a priori knowledge is inevitable. In case that the weak link hypothesis [180] that many weak links are crucial for a correct network modelling is true, then even the use of sparsity assumptions in the reconstruction of the networks are at least questionable. Therefore methods for a reliable verification of the reconstructed networks are highly desired.

Another possibility in the future would be to use hypercomputation methods [181], which are based on the principal of least action [21] and therefore are able to solve NP-complete problems in polynomial time. It's assumed that the cell is somehow exactly doing this inherently.

Since there are already quantum algorithm for learning of Boolean juntas available [182], it's also possible to use quantum computers for the network inference as soon as efficient and generally programmable quantum computers are available.

Last but not least, it should be mentioned that the simulation of biomolecular networks is also important in synthetic biology, where the goal is to engineer and optimize organisms for the industrial biological-based production of materials like e.g. bio-fuel, biopharmaceuticals. *Nguyen et.al.* for instance describe a converter, which can convert SBML models into Synthetic Biology Open Language (SBOL) [183] and *Kobayashi and Hiraishi* describe the design of Boolean networks with a desired attractor structure [184].

**Conflict of interest:** All authors declare that they have no conflicts of interest.

**Acknowledgements / Funding:**

GM was funded by the de.NBI project (http://www.denbi.de) of the German Federal Ministry of Education and research (BMBF) [grant number FKZ 031 A 534A].

**References:**

1. Machado D, Costa RS, Rocha M, Ferreira EC, Tidor B, Rocha I. Modeling formalisms in Systems Biology. *AMB Express*, 1, 45 (2011).

2. Cheng DZ, Qi HS, Xue AC. A survey on semi-tensor product of matrices. *J Syst Sci Complex*, 20(2), 304-322 (2007).

3. Laubenbacher R, Jarrah AS. Algebraic Models of Biochemical Networks. 467, 163-196 (2009).

4. Tran QN. Algebraic model checking for Boolean gene regulatory networks. *Advances in experimental medicine and biology*, 696, 113-122 (2011).

5. Fisher J, Henzinger TA. Executable cell biology. *Nature biotechnology*, 25(11), 1239-1249 (2007).

6. Karlebach G, Shamir R. Modelling and analysis of gene regulatory networks. *Nature reviews. Molecular cell biology*, 9(10), 770-780 (2008).

7. Materi W, Wishart DS. Computational systems biology in drug discovery and development: methods and applications. *Drug discovery today*, 12(7-8), 295-303 (2007).

8. Vijesh N, Chakrabarti SK, Sreekumar J. Modeling of gene regulatory networks: A review. *Journal of Biomedical Science and Engineering*, 06(02), 223-231 (2013).

9. Bartocci E, Lió P. Computational Modeling, Formal Analysis, and Tools for Systems Biology. *PLoS Comput Biol*, 12(1), e1004591 (2016).

10. Najafi A, Bidkhori G, Bozorgmehr JH, Koch I, Masoudi-Nejad A. Genome scale modeling in systems biology: algorithms and resources. *Current genomics*, 15(2), 130-159 (2014).

11. Murrugarra D, Veliz-Cuba A, Aguilar B, Arat S, Laubenbacher R. Modeling stochasticity and variability in gene regulatory networks. *EURASIP journal on bioinformatics & systems biology*, 2012(1), 5 (2012).

12. Schivo S, Scholma J, van der Vet PE *et al.* Modelling with ANIMO: between fuzzy logic and differential equations. *BMC Syst Biol*, 10(1), 56 (2016).

13. Albert R. Boolean Modeling of Genetic Regulatory Networks. *Lect. Notes Phys.*, 650, 459-481 (2004).

14. Saadatpour A, Albert R. Boolean modeling of biological regulatory networks: a methodology tutorial. *Methods*, 62(1), 3-12 (2013).

15. Wang RS, Saadatpour A, Albert R. Boolean modeling in systems biology: an overview of methodology and applications. *Physical biology*, 9(5) (2012).

16. Chaouiya C, Remy E, Thieffry D. Petri net modelling of biological regulatory networks. *Journal of Discrete Algorithms*, 6(2), 165-177 (2008).

17. Koch I. *Modeling in Systems Biology: The Petri Net Approach* (Springer, 2010).

18. Brickenstein M, Dreyer A. POLYBORI: A framework for Grobner-basis computations with Boolean polynomials. *J Symb Comput*, 44(9), 1326-1345 (2009).

19. Ilea M, Turnea M, Rotariu M. Ordinary differential equations with applications in molecular biology. *Revista medico-chirurgicala a Societatii de Medici si Naturalisti din Iasi*, 116(1), 347-352 (2012).

20. Stigler B. Polynomial dynamical systems in systems biology. *Proc Sym Ap*, 64, 53-84 (2007).

21. Mainzer K, Chua, L. *Local Activity Principle: The Cause of Complexity and Symmetry Breaking* (Imperial College Press, 2013).

22. Cowan NJ, Chastain EJ, Vilhena DA, Freudenberg JS, Bergstrom CT. Nodal dynamics, not degree distributions, determine the structural controllability of complex networks. *PloS one*, 7(6), e38398 (2012).

23. Stetter HJ. *Numerical Polynomial Algebra* (Society for Industrial and Applied Mathematics, 2004).

24. Alterovitz G, Liu J, Afkhami E, Ramoni MF. Bayesian methods for proteomics. *Proteomics*, 7(16), 2843-2855 (2007).

25. Dojer N, Gambin A, Mizera A, Wilczynski B, Tiuryn J. Applying dynamic Bayesian networks to perturbed gene expression data. *BMC bioinformatics*, 7, 249 (2006).

26. Ciocchetta F, Hillston J. Bio-PEPA: A framework for the modelling and analysis of biological systems. *Theoretical Computer Science*, 410(33-34), 3065-3084 (2009).

27. Deutsch A. *Cellular Automaton Modeling of Biological Pattern Formation: Characterization, Applications, and Analysis* (Birkhäuser, 2004).

28. Wishart DS, Yang R, Arndt D, Tang P, Cruz J. Dynamic cellular automata: an alternative approach to cellular simulation. *In silico biology*, 5(2), 139-161 (2005).

29. Berenguier D, Chaouiya C, Monteiro PT *et al.* Dynamical modeling and analysis of large cellular regulatory networks. *Chaos*, 23(2), 025114 (2013).

30. Ruklisa D, Brazma A, Viksna J. Reconstruction of gene regulatory networks under the finite state linear model. *Genome informatics. International Conference on Genome Informatics*, 16(2), 225-236 (2005).

31. Hinkelmann F, Murrugarra D, Jarrah AS, Laubenbacher R. A mathematical framework for agent based models of complex biological networks. *Bull Math Biol*, 73(7), 1583-1602 (2011).

32. Chylek LA, Harris LA, Tung CS, Faeder JR, Lopez CF, Hlavacek WS. Rule-based modeling: a computational approach for studying biomolecular site dynamics in cell signaling systems. *Wiley interdisciplinary reviews. Systems biology and medicine*, 6(1), 13-36 (2014).

33. De Jong H, Gouze JL, Hernandez C, Page M, Sari T, Geiselmann J. Qualitative simulation of genetic regulatory networks using piecewise-linear models. *B Math Biol*, 66(2), 301-340 (2004).

34. Gillespie DT. Stochastic simulation of chemical kinetics. *Annu Rev Phys Chem*, 58, 35-55 (2007).

35. Klann MT, Lapin A, Reuss M. Stochastic simulation of signal transduction: impact of the cellular architecture on diffusion. *Biophysical journal*, 96(12), 5122-5129 (2009).

36. Ribeiro AS. Stochastic and delayed stochastic models of gene expression and regulation. *Mathematical biosciences*, 223(1), 1-11 (2010).

37. Szekely T, Jr., Burrage K. Stochastic simulation in systems biology. *Computational and structural biotechnology journal*, 12(20-21), 14-25 (2014).

38. Wu FX. Gene Regulatory Network modelling: a state-space approach. *International journal of data mining and bioinformatics*, 2(1), 1-14 (2008).

39. Noor AS, E.; Nounou, M.; Nounou, H. INFERRING GENE REGULATORY NETWORKS WITH NONLINEAR MODELS VIA EXPLOITING SPARSITY‡. In: *ICASSP 2012.* (Ed.^(Eds) (IEEE, 2012)

40. G P. Computing with Membranes: Attacking NP-Complete Problems. In: *Unconventional Models of Computation, UMC'2K* (Ed.^(Eds) (Springer, 2001) 94-115.

41. Manca V. From P to MP Systems. In: *WMC 2009 : Membrane Computing.* G., P, M.J., P-J, A, R-N, G, R, A., S (Ed.^(Eds) (Springer, 2009) 74-94.

42. Raue A, Kreutz C, Maiwald T, Klingmuller U, Timmer J. Addressing parameter identifiability by model-based experimentation. *IET systems biology*, 5(2), 120-130 (2011).

43. Dougherty ER, Shmulevich I. On the limitations of biological knowledge. *Current genomics*, 13(7), 574-587 (2012).

44. Raue A, Schilling M, Bachmann J *et al.* Lessons learned from quantitative dynamical modeling in systems biology. *PloS one*, 8(9), e74335 (2013).

45. Chindelevitch L, Trigg J, Regev A, Berger B. An exact arithmetic toolbox for a consistent and reproducible structural analysis of metabolic network models. *Nature communications*, 5, 4893 (2014).

46. Gutenkunst RN, Waterfall JJ, Casey FP, Brown KS, Myers CR, Sethna JP. Universally sloppy parameter sensitivities in systems biology models. *PLoS Comput Biol*, 3(10), 1871-1878 (2007).

47. Wynn ML, Consul N, Merajver SD, Schnell S. Logic-based models in systems biology: a predictive and parameter-free network analysis method. *Integrative biology : quantitative biosciences from nano to macro*, 4(11), 1323-1337 (2012).

48.    Davidich M, Bornholdt S. The transition from differential equations to Boolean networks: a case study in simplifying a regulatory network model. *Journal of theoretical biology*, 255(3), 269-277 (2008).

49.    Robeva R, Laubenbacher R. Mathematical biology education: beyond calculus. *Science*, 325(5940), 542-543 (2009).

50.    Kauffman SA. Metabolic stability and epigenesis in randomly constructed genetic nets. *Journal of theoretical biology*, 22(3), 437-467 (1969).

51.    Kauffman SA. *The Origins of Order: Self-Organization and Selection in Evolution* (Oxford University Press, 1993).

52.    Gershenson C. Classification of random boolean networks. In: *Artificial Life VIII: Proceedings of the Eight International Conference on Artificial Life.* (MIT Press, 2002) 1-8.

53.    Gershenson C, Broekaert J, Aerts D. Contextual random boolean networks. *Lect Notes Artif Int*, 2801, 615-624 (2003).

54.    Zhu XW, Gerstein M, Snyder M. Getting connected: analysis and principles of biological networks. *Gene Dev*, 21(9), 1010-1024 (2007).

55.    Lima-Mendez G, van Helden J. The powerful law of the power law and other myths in network biology. *Molecular bioSystems*, 5(12), 1482-1493 (2009).

56.    Mayer G, Marcus K, Eisenacher M, Kohl M. Boolean modeling techniques for protein co-expression networks in systems medicine. *Expert Rev Proteomics*, 13(6), 555-569 (2016).

57.    Lu J, Zhong J, Tang Y, Huang T, Cao J, Kurths J. Synchronization in output-coupled temporal Boolean networks. *Scientific reports*, 4, 6292 (2014).

58.    Zhong J, Lu J, Huang T, Cao J. Synchronization of master–slave Boolean networks with impulsive effects: Necessary and sufficient criteria. *Neurocomputing*, 143, 269-274 (2014).

59.    Lu J, Zhong J, Li L, Ho DW, Cao J. Synchronization Analysis of Master-Slave Probabilistic Boolean Networks. *Sci Rep*, 5, 13437 (2015).

60.    Zou C, Wei X, Zhang Q, Zhou C. Passivity of Reaction–Diffusion Genetic Regulatory  Networks with Time-Varying Delays    *10.1007/s11063-017-9682-7*,    (2017).

61.    Bornholdt S. Boolean network models of cellular regulation: prospects and limitations. *Journal of the Royal Society, Interface / the Royal Society*, 5 Suppl 1, S85-94 (2008).

62.    Xiao Y. A tutorial on analysis and simulation of boolean gene regulatory network models. *Current genomics*, 10(7), 511-525 (2009).

63.    Benso A, Di Carlo S, Politano G, Savino A, Vasciaveo A. An extended gene protein/products Boolean network model including post-transcriptional regulation. *Theoretical biology & medical modelling*, 11 Suppl 1, S5 (2014).

64.    Simak M, Yeang CH, Lu HH. Exploring candidate biological functions by Boolean Function Networks for Saccharomyces cerevisiae. *PLoS One*, 12(10), e0185475 (2017).

65.    Song MJ, Lewis CK, Lance ER *et al.* Reconstructing generalized logical networks of transcriptional regulation in mouse brain from temporal gene expression data. *EURASIP journal on bioinformatics & systems biology*, 545176 (2009).

66.    Al Qazlan TA, Hamdi-Cherif A, Kara-Mohamed C. State of the art of fuzzy methods for gene regulatory networks inference. *TheScientificWorldJournal*, 2015, 148010 (2015).

67.    Morris MK, Saez-Rodriguez J, Clarke DC, Sorger PK, Lauffenburger DA. Training Signaling Pathway Maps to Biochemical Data with Constrained Fuzzy Logic: Quantitative Analysis of Liver Cell Responses to Inflammatory Stimuli. *Plos Computational Biology*, 7(3) (2011).

68.    Jarrah AS, Raposa B, Laubenbacher R. Nested canalyzing, unate cascade, and polynomial functions. *Physica D*, 233(2), 167-174 (2007).

69.    Veliz-Cuba A, Aguilar B, Laubenbacher R. Dimension Reduction of Large Sparse AND-NOT Network Models. *Electronic Notes in Theoretical Computer Science*, 316, 83-95 (2015).

70. Veliz-Cuba A, Buschur K, Hamershock R, Kniss A, Wolff E, Laubenbacher R. AND-NOT logic framework for steady state analysis of Boolean network models. *Appl Math Inform Sci*, 7(4), 1263-1274 (2013).

71. Shmulevich I, Dougherty ER, Mang W. From Boolean to probabilistic Boolean networks as models of genetic regulatory networks. *P Ieee*, 90(11), 1778-1792 (2002).

72. Trairatphisan P, Mizera A, Pang J, Tantar AA, Schneider J, Sauter T. Recent development and biomedical applications of probabilistic Boolean networks. *Cell communication and signaling : CCS*, 11, 46 (2013).

73. Liang J, Han J. Stochastic Boolean networks: an efficient approach to modeling gene regulatory networks. *BMC systems biology*, 6, 113 (2012).

74. Higa CH, Louzada VH, Andrade TP, Hashimoto RF. Constraint-based analysis of gene interactions using restricted boolean networks and time-series data. *BMC Proc*, 5 Suppl 2, S5 (2011).

75. Tran V, McCall MN, McMurray HR, Almudevar A. On the underlying assumptions of threshold Boolean networks as a model for genetic regulatory network behavior. *Frontiers in genetics*, 4, 263 (2013).

76. Brouard C, Vrain C, Dubois J, Castel D, Debily MA, d'Alche-Buc F. Learning a Markov Logic network for supervised gene regulatory network inference. *BMC bioinformatics*, 14, 273 (2013).

77. Dingel J, Milenkovic O. List-decoding methods for inferring polynomials in finite dynamical gene network models. *Bioinformatics*, 25(13), 1686-1693 (2009).

78. Lichtenberg GE, A. Multilinear Algebraic Boolean Modelling with Tensor Decompositions Techniques. In: *IFAC World Congress.* (Ed.^(Eds) (Milano, 2011)

79. Brickenstein M, Dreyer A. Grobner-free normal forms for Boolean polynomials. *J Symb Comput*, 48, 37-53 (2013).

80. Faisal S, Lichtenberg G, Trump S, Attinger S. Structural properties of continuous representations of Boolean functions for gene network modelling. *Automatica*, 46(12), 2047-2052 (2010).

81. Hasegawa T, Yamaguchi R, Nagasaki M, Miyano S, Imoto S. Inference of gene regulatory networks incorporating multi-source biological knowledge via a state space model with L1 regularization. *PloS one*, 9(8), e105942 (2014).

82. Grigoriev A. On the number of protein-protein interactions in the yeast proteome. *Nucleic acids research*, 31(14), 4157-4161 (2003).

83. Guelzim N, Bottani S, Bourgine P, Kepes F. Topological and causal structure of the yeast transcriptional regulatory network. *Nature genetics*, 31(1), 60-63 (2002).

84. Jarrah AS, Laubenbacher R, Veliz-Cuba A. The dynamics of conjunctive and disjunctive Boolean network models. *Bull Math Biol*, 72(6), 1425-1447 (2010).

85. Steggles LJ, Banks R, Shaw O, Wipat A. Qualitatively modelling and analysing genetic regulatory networks: a Petri net approach. *Bioinformatics*, 23(3), 336-343 (2007).

86. Gianchandani EP, Papin JA, Price ND, Joyce AR, Palsson BO. Matrix formalism to describe functional states of transcriptional regulatory systems. *PLoS Comput Biol*, 2(8), e101 (2006).

87. Wittmann DM, Krumsiek J, Saez-Rodriguez J, Lauffenburger DA, Klamt S, Theis FJ. Transforming Boolean models to continuous models: methodology and application to T-cell receptor signaling. *BMC systems biology*, 3, 98 (2009).

88. Smallbone K, Simeonidis E, Broomhead DS, Kell DB. Something from nothing - bridging the gap between constraint-based and kinetic modelling. *Febs J*, 274(21), 5576-5585 (2007).

89. Gilbert DH, M. From Petri nets to differential equations An integrated approach for biochemical network analysis. In: *Petri Nets and Other Models of Concurrency - ICATPN 2006.* (2006) 181-200.

90. Egri-Nagy A, Nehaniv CL. Algebraic properties of automata associated to Petri nets and applications to computation in biological systems. *Biosystems*, 94(1-2), 135-144 (2008).

91. Calder MG, S.; Hillston, J. Automatically deriving ODEs from process algebra models of signalling pathways. In: *Computational Methods in Systems Bi9ology.* (Springer, 2005) 204-215.

92.     Feret J, Danos V, Krivine J, Harmer R, Fontana W. Internal coarse-graining of molecular systems. *Proceedings of the National Academy of Sciences of the United States of America*, 106(16), 6453-6458 (2009).

93.     Pedersen M, Plotkin GD. A Language for Biochemical Systems: Design and Formal Specification. *Lect N Bioinformat*, 5945, 77-145 (2010).

94.     Diener FD, A.; Bernot, G.; Comet, J.P.; Eyssette, F. Correspondence between discrete and piecewise linear models of gene regulatory networks.     (2013).

95.     Lahdesmaki H, Hautaniemi S, Shmulevich I, Yli-Harja O. Relationships between probabilistic Boolean networks and dynamic Bayesian networks as models of gene regulatory networks. *Signal Process*, 86(4), 814-834 (2006).

96.     Laubenbacher R, Stigler B. A computational algebra approach to the reverse engineering of gene regulatory networks. *Journal of theoretical biology*, 229(4), 523-537 (2004).

97.     Dimitrova E, Garcia-Puente LD, Hinkelmann F *et al.* Parameter estimation for Boolean models of biological networks. *Theoretical Computer Science*, 412(26), 2816-2826 (2011).

98.     Veliz-Cuba A, Jarrah AS, Laubenbacher R. Polynomial algebra of discrete models in systems biology. *Bioinformatics*, 26(13), 1637-1643 (2010).

99.     Castellini A, Manca V, Marchetti L. MP Systems and Hybrid Petri Nets. 129, 53-62 (2008).

100.    Cheng D-Z, Qi HPD, Li ZMS. *Analysis and control of boolean networks : a semi-tensor product approach* (Springer, London, 2011).

101.    Jianquan L, Haitao L, Yang L, Fangfei L. Survey on semi-tensor product method with its applications in logical networks and other finite-valued systems. *IET Control Theory & Applications*  11(13), 2040-2047 (2017).

102.    Cheng DZ, Qi HS. A Linear Representation of Dynamics of Boolean Networks. *Ieee Transactions on Automatic Control*, 55(10), 2251-2258 (2010).

103.    Vera-Licona P, Jarrah A, Garcia-Puente LD, McGee J, Laubenbacher R. An algebra-based method for inferring gene regulatory networks. *BMC systems biology*, 8, 37 (2014).

104.    Luo C, Wang XY. Algebraic Representation of Asynchronous Multiple-Valued Networks and Its Dynamics. *Ieee Acm T Comput Bi*, 10(4), 927-938 (2013).

105.    Zhan J, Lu S, Yang G. Analysis of Boolean Networks Using an Optimized Algorithm of Structure Matrix Based on Semi-tensor Product. *Journal of Computers*, 8(6) (2013).

106.    Cheng D, Zhao Y. Identification of Boolean control networks. *Automatica*, 47(4), 702-710 (2011).

107.    Cheng DZ, Zhao Y, Liu JB. Optimal Control of Finite-Valued Networks. *Asian J Control*, 16(4), 1179-1190 (2014).

108.    Zou YL, Zhu JD. Kalman decomposition for Boolean control networks. *Automatica*, 54, 65-71 (2015).

109.    Liu Y, Chen HW, Wu B, Sun LJ. A Mayer-type optimal control for multivalued logic control networks with undesirable states. *Appl Math Model*, 39(12), 3357-3365 (2015).

110.    Mou C. Solving Polynomial Systems over Finite Fields: Algorithms, Implementation and Applications. (Ed.^(Eds) (Université Pierre et Marie Curie, 2013)

111.    Veliz-Cuba A, Aguilar B, Hinkelmann F, Laubenbacher R. Steady state analysis of Boolean molecular network models via model reduction and computational algebra. *BMC bioinformatics*, 15, 221 (2014).

112.    Faisal S, Lichtenberg G, Werner H. Polynomial models of gene dynamics. *Neurocomputing*, 71(13-15), 2711-2719 (2008).

113.    Karl S, Dandekar T. Jimena: efficient computing and system state identification for genetic regulatory networks. *BMC bioinformatics*, 14, 306 (2013).

114.    Schaub MA, Henzinger TA, Fisher J. Qualitative networks: a symbolic approach to analyze biological signaling networks. *BMC systems biology*, 1, 4 (2007).

115.     Garg A, Xenarios I, Mendoza L, DeMicheli G. An efficient method for dynamic analysis of gene regulatory networks and in silico gene perturbation experiments. *Research in Computational Molecular Biology, Proceedings*, 4453, 62-76 (2007).

116.     Klarner H, Bockmayr A, Siebert H. Computing Symbolic Steady States of Boolean Networks. *Cellular Automata: 11th International Conference on Cellular Automata for Research and Industry*, 8751, 561-570 (2014).

117.     Klarner H, Siebert H. Approximating Attractors of Boolean Networks by Iterative CTL Model Checking. *Frontiers in bioengineering and biotechnology*, 3, 130 (2015).

118.     Langmead CJ, Jha SK. Symbolic approaches for finding control strategies in Boolean Networks. *Journal of bioinformatics and computational biology*, 7(2), 323-338 (2009).

119.     Kobayashi K, Hiraishi K. Verification and Optimal Control of Context-Sensitive Probabilistic Boolean Networks Using Model Checking and Polynomial Optimization. *Sci World J*,    (2014).

120.     Wilkinson MD, Dumontier M, Aalbersberg IJ *et al.* The FAIR Guiding Principles for scientific data management and stewardship. *Sci Data*, 3, 160018 (2016).

121.     Drager A, Palsson BO. Improving collaboration by standardization efforts in systems biology. *Frontiers in bioengineering and biotechnology*, 2, 61 (2014).

122.     Le Novere N, Finney A, Hucka M *et al.* Minimum information requested in the annotation of biochemical models (MIRIAM). *Nature biotechnology*, 23(12), 1509-1515 (2005).

123.     Waltemath D, Adams R, Beard DA *et al.* Minimum Information About a Simulation Experiment (MIASE). *Plos Computational Biology*, 7(4) (2011).

124.     Hucka M, Nickerson DP, Bader GD *et al.* Promoting Coordinated Development of Community-Based Information Standards for Modeling in Biology: The COMBINE Initiative. *Front Bioeng Biotechnol*, 3, 19 (2015).

125.     Juty N, Le Novere N, Laibe C. Identifiers.org and MIRIAM Registry: community resources to provide persistent identification. *Nucleic acids research*, 40(Database issue), D580-586 (2012).

126.     Bergmann FT, Adams R, Moodie S *et al.* COMBINE archive and OMEX format: one file to share all information to reproduce a modeling project. *BMC bioinformatics*, 15 (2014).

127.     Naldi A, Monteiro PT, Mussel C *et al.* Cooperative development of logical modelling standards and tools with CoLoMoTo. *Bioinformatics*, 31(7), 1154-1159 (2015).

128.     Chaouiya C, Berenguier D, Keating SM *et al.* SBML qualitative models: a model representation format and infrastructure to foster interactions between qualitative modelling formalisms and tools. *BMC systems biology*, 7 (2013).

129.     Hoehndorf R, Schofield PN, Gkoutos GV. The role of ontologies in biological and biomedical research: a functional perspective. *Briefings in bioinformatics*,    (2015).

130.     Lister AL, P.; Pocock, M.; Wipat, A. Annotation of SBML models through rule-based semantic integration. *Journal of Biomedical Semantics*, 1(Suppl 1):S3 (2010).

131.     Mayer G, Jones AR, Binz PA *et al.* Controlled vocabularies and ontologies in proteomics: overview, principles and practice. *Biochimica et biophysica acta*, 1844(1 Pt A), 98-107 (2014).

132.     Courtot M, Juty N, Knupfer C *et al.* Controlled vocabularies and semantics in systems biology. *Molecular systems biology*, 7, 543 (2011).

133.     Saez-Rodriguez J, Goldsipe A, Muhlich J *et al.* Flexible informatics for linking experimental data to mathematical models via DataRail. *Bioinformatics*, 24(6), 840-847 (2008).

134.     Hucka M, Finney A, Sauro HM *et al.* The systems biology markup language (SBML): a medium for representation and exchange of biochemical network models. *Bioinformatics*, 19(4), 524-531 (2003).

135.     Garny A, Nickerson DP, Cooper J *et al.* CellML and associated tools and techniques. *Philosophical transactions. Series A, Mathematical, physical, and engineering sciences*, 366(1878), 3017-3043 (2008).

136. Beltrame L, Calura E, Popovici RR *et al.* The Biological Connection Markup Language: a SBGN-compliant format for visualization, filtering and analysis of biological pathways. *Bioinformatics*, 27(15), 2127-2133 (2011).

137. Kyoda K, Tohsato Y, Ho KH, Onami S. Biological Dynamics Markup Language (BDML): an open format for representing quantitative biological dynamics data. *Bioinformatics*, 31(7), 1044-1052 (2015).

138. Muggianu F, Benso A, Bardini R, Hu E, Politano G, Di Carlo S. Modeling biological complexity using Biology System Description Language (BiSDL). In: *IEEE International Conference on Bioinformatics and Biomedicine (BIBM)* (Ed.^(Eds) (2018) 713-717.

139. Waltemath D, Adams R, Bergmann FT *et al.* Reproducible computational biology experiments with SED-ML--the Simulation Experiment Description Markup Language. *BMC systems biology*, 5, 198 (2011).

140. Dada JO, Spasic I, Paton NW, Mendes P. SBRML: a markup language for associating systems biology data with models. *Bioinformatics*, 26(7), 932-938 (2010).

141. Catlett NL, Bargnesi AJ, Ungerer S *et al.* Reverse causal reasoning: applying qualitative causal knowledge to the interpretation of high-throughput data. *BMC bioinformatics*, 14, 340 (2013).

142. Slater T. Recent advances in modeling languages for pathway maps and computable biological networks. *Drug discovery today*, 19(2), 193-198 (2014).

143. Kutmon M, van Iersel MP, Bohler A *et al.* PathVisio 3: an extendable pathway analysis toolbox. *PLoS Comput Biol*, 11(2), e1004085 (2015).

144. Demir E, Cary MP, Paley S *et al.* The BioPAX community standard for pathway data sharing. *Nature biotechnology*, 28(9), 935-942 (2010).

145. van Iersel MP, Villeger AC, Czauderna T *et al.* Software support for SBGN maps: SBGN-ML and LibSBGN. *Bioinformatics*, 28(15), 2016-2021 (2012).

146. Gauges R, Rost U, Sahle S, Wegner K. A model diagram layout extension for SBML. *Bioinformatics*, 22(15), 1879-1885 (2006).

147. Le Novere N, Hucka M, Mi H *et al.* The Systems Biology Graphical Notation. *Nature biotechnology*, 27(8), 735-741 (2009).

148. Malone J, Holloway E, Adamusiak T *et al.* Modeling sample variables with an Experimental Factor Ontology. *Bioinformatics*, 26(8), 1112-1118 (2010).

149. Saez-Rodriguez J, Alexopoulos LG, Stolovitzky G. Setting the standards for signal transduction research. *Sci Signal*, 4(160), pe10 (2011).

150. Keller R, Dorr A, Tabira A *et al.* The systems biology simulation core algorithm. *BMC systems biology*, 7, 55 (2013).

151. Rodriguez N, Thomas A, Watanabe L *et al.* JSBML 1.0: providing a smorgasbord of options to encode systems biology models. *Bioinformatics*, 31(20), 3383-3386 (2015).

152. Somogyi ET, Bouteiller JM, Glazier JA *et al.* libRoadRunner: a high performance SBML simulation and analysis library. *Bioinformatics*, 31(20), 3315-3321 (2015).

153. Ezzelle J, Rodriguez-Chavez IR, Darden JM *et al.* Guidelines on good clinical laboratory practice: Bridging operations between research and clinical research laboratories. *J Pharmaceut Biomed*, 46(1), 18-29 (2008).

154. Manolio TA, Chisholm RL, Ozenberger B *et al.* Implementing genomic medicine in the clinic: the future is here. *Genetics in medicine : official journal of the American College of Medical Genetics*, 15(4), 258-267 (2013).

155. Cox J, Matic I, Hilger M *et al.* A practical guide to the MaxQuant computational platform for SILAC-based quantitative proteomics. *Nat Protoc*, 4(5), 698-705 (2009).

156. Griss J, Jones AR, Sachsenberg T *et al.* The mzTab data exchange format: communicating mass-spectrometry-based proteomics and metabolomics experimental results to a wider audience. *Molecular & cellular proteomics : MCP*, 13(10), 2765-2775 (2014).

157. Vizcaino JA, Csordas A, Del-Toro N *et al.* 2016 update of the PRIDE database and its related tools. *Nucleic acids research*, 44 (2015).

158. Walzer M, Qi D, Mayer G *et al.* The mzQuantML data standard for mass spectrometry-based quantitative studies in proteomics. *Molecular & cellular proteomics : MCP*, 12(8), 2332-2340 (2013).

159. Vizcaino JA, Deutsch EW, Wang R *et al.* ProteomeXchange provides globally coordinated proteomics data submission and dissemination. *Nature biotechnology*, 32(3), 223-226 (2014).

160. A. C, F. F, S. S. Probably Approximately Correct Learning of Regulatory Networks from Time-Series Data. In: *CMSB - 15th International Conference on Computational Methods for Systems Biology.* J., F, H., K (Ed.^(Eds) (Springer, Darmstadt, 2017) 74-90.

161. Valiant L. ***Probably Approximately Correct: Nature's Algorithms for Learning and Prospering in a Complex World*** (Basic Books, New York, 2014).

162. Mitsos A, Melas IN, Morris MK, Saez-Rodriguez J, Lauffenburger DA, Alexopoulos LG. Non Linear Programming (NLP) formulation for quantitative modeling of protein signal transduction pathways. *PLoS One*, 7(11), e50085 (2012).

163. Knijnenburg TA, Klau GW, Iorio F *et al.* Logic models to predict continuous outputs based on binary inputs with an application to personalized cancer therapy. *Sci Rep*, 6, 36812 (2016).

164. Dorier J, Crespo I, Niknejad A, Liechti R, Ebeling M, Xenarios I. Boolean regulatory network reconstruction using literature based knowledge with a genetic algorithm optimization method. *BMC Bioinformatics*, 17(1), 410 (2016).

165. Chang YH, Gray JW, Tomlin CJ. Exact reconstruction of gene regulatory networks using compressive sensing. *BMC Bioinformatics*, 15, 400 (2014).

166. Parisi F, Koeppl H, Naef F. Network inference by combining biologically motivated regulatory constraints with penalized regression. *Ann N Y Acad Sci*, 1158, 114-124 (2009).

167. Miryala SK, Anbarasu A, Ramaiah S. Discerning molecular interactions: A comprehensive review on biomolecular interaction databases and network analysis tools. *Gene*, 642, 84-94 (2018).

168. Wu X, Hasan MA, Chen JY. Pathway and network analysis in proteomics. *J Theor Biol*, 362, 44-52 (2014).

169. Fabregat A, Korninger F, Viteri G *et al.* Reactome graph database: Efficient access to complex pathway data. *PLoS Comput Biol*, 14(1), e1005968 (2018).

170. Slenter DN, Kutmon M, Hanspers K *et al.* WikiPathways: a multifaceted pathway database bridging metabolomics to other omics research. *Nucleic Acids Res*, 46(D1), D661-D667 (2018).

171. Vasaikar SV, Padhi AK, Jayaram B, Gomes J. NeuroDNet - an open source platform for constructing and analyzing neurodegenerative disease networks. *BMC Neurosci*, 14, 3 (2013).

172. Meng X, Wang J, Yuan C *et al.* CancerNet: a database for decoding multilevel molecular interactions across diverse cancer types. *Oncogenesis*, 4, e177 (2015).

173. Crespo I, Krishna A, Le Béchec A, del Sol A. Predicting missing expression values in gene regulatory networks using a discrete logic modeling optimization guided by network stable states. *Nucleic Acids Res*, 41(1), e8 (2013).

174. Rodriguez A, Crespo I, Androsova G, del Sol A. Discrete Logic Modelling Optimization to Contextualize Prior Knowledge Networks Using PRUNET. *PLoS One*, 10(6), e0127216 (2015).

175. Gómez-Vela F, Díaz-Díaz N. Gene network biological validity based on gene-gene interaction relevance. *ScientificWorldJournal*, 2014, 540679 (2014).

176. Liu Y-Y, Barabási A-L. Control principles of complex systems. *REVIEWS OF MODERN PHYSICS*, 88(3), 1-58 (2016).

177. Schwab JD, Kestler HA. Automatic Screening for Perturbations in Boolean Networks. *Front Physiol*, 9, 431 (2018).

178. Li R, Yang M, Chu T. Controllability and observability of Boolean networks arising from biology. *Chaos*, 25(2), 023104 (2015).

179. Miskov-Zivanov N, Bresticker A, Krishnaswamy D *et al.* Regulatory network analysis acceleration with reconfigurable hardware. *Conf Proc IEEE Eng Med Biol Soc*, 2011, 149-152 (2011).

180. Csermely P. *Weak Links: The Universal Key to the Stability of Networks and Complex Systems* (Springer, 2009).

181. MALDONADO CE, GÓMEZ CRUZ NA. Biological Hypercomputation: A New Research Problem in Complexity Theory *Complexity*, 8-18 (2014 ).

182. Atıcı A, Servedio RA. Quantum Algorithms for Learning and Testing Juntas. *Quantum Information Processing*, 6(5), 323-348 (2007).

183. Nguyen T, Roehner N, Zundel Z, Myers CJ. A Converter from the Systems Biology Markup Language to the Synthetic Biology Open Language. *ACS Synth Biol*, 5(6), 479-486 (2016).

184. Kobayashi K, Hiraishi K. ILP/SMT-Based Method for Design of Boolean Networks Based on Singleton Attractors. *IEEE/ACM Trans Comput Biol Bioinform*, 11(6), 1253-1259 (2014).

Supplementary table 1: Overview about software packages and libraries supporting the logical modelling and analysis approaches.

| Name | Short description | URL (Last accessed 29th February 2020) | Reference |
|---|---|---|---|
| ADAM | Analysis of Dynamic Algebraic Models. | http://adam.plantsimlab.org/adam.html | [1] |
| ASSA-PBN | Modelling, simulation and analysis of Probabilistic Boolean Networks. | http://satoss.uni.lu/software/ASSA-PBN/ | [2] |
| ASP-G | Answer Set Programming method for finding attractors in Boolean networks. | http://bioinformatics.intec.ugent.be/kmarchal/Supplementary_Information_Musthofa_2014/asp-g.zip | [3] |
| ATLANTIS | Attractor Landscape Analysis Toolbox for Cell Fate Discovery and Reprogramming. | https://github.com/BIRL/ATLANTIS | [4] |
| bioLQM | Java library for Logical Qualitative Models of biological networks. | https://github.com/colomoto/bioLQM | [5] |
| BiTrinA | R package for the binarization and trinarization of expression profile data. | https://cran.r-project.org/web/packages/BiTrinA/index.html | [6] |
| BN++ (BNPP) / BiNA | C++ Library for modelling and analyzing biochemical networks. | https://sourceforge.net/projects/bnpp/ | [7] |
| BMA | Bio Model Analyzer, a visual tool for modelling and analyzing biological networks. | http://biomodelanalyzer.research.microsoft.com/tool.html | [8] |
| B-NEM | R package for Boolean Nested Effect Models (B-NEM). | https://github.com/MartinFXP/B-NEM | [9,10] |
| BNS | An analysis tool for the computation of attractors in Boolean network models with synchronous update. | https://people.kth.se/~dubrova/bns.html | [11] |

| BNT / EBNT | (Extended) Boolean Network Toolkit: A C++ toolkit for the computing the attractors and the state space of Boolean network models. | http://www.sysbio.polito.it/index.php/tools-and-downloads/item/208-boolean-regulatory-network-simulator | --- |
|---|---|---|---|
| BooleanDynamicModeling | Java library for the dynamic modeling of Boolean networks. | https://github.com/jgtz/BooleanDynamicModeling/ | [12] |
| BooleanNet | Python program for simulation of biological regulatory networks as Boolean networks. | https://github.com/ialbert/booleannet<br>http://atlas.bx.psu.edu/booleannet/booleannet.html | [13] |
| Boolean Network Toolkit | C++ library for simulation, attractor sampling and Derrida plot computation of large Boolean networks. | http://sourceforge.net/projects/booleannetwork/ | [14] |
| BooleSim | BooleSim is a simulation program for Boolean networks, running in a browser. | http://rumo.biologie.hu-berlin.de/boolesim/<br>https://github.com/matthiasbock/BooleSim | [15] |
| BoolFilter | R package for state estimation as well as network inference of Partially-Observed Boolean Dynamical Systems. | https://CRAN.R-project.org/package=BoolFilter | [16] |
| BoolNet | R package for the construction, simulation and analysis of Boolean network models. | https://cran.r-project.org/web/packages/BoolNet/ | [17] |
| BTR | Tools for inference and analysis of asynchronous Boolean nets | https://rdrr.io/cran/BTR/ | [18] |
| CABeRNET | A Cytoscape [19] plugin for the simulation and analysis of Boolean models. | http://bimib.disco.unimib.it/index.php/CABERNET | [20,21] |

| CANA | Python package for control and canalyzing in Boolean networks. | https://pypi.org/project/cana/ | [22] |
|------|------|------|------|
| caspo | Reasoning on the response of logical signaling networks with Answer Set Programming | http://bioasp.github.io/caspo/ | [23,24] |
| Cell Collective | A web-based platform for the construction, simulation, and analysis of Boolean-based models. | http://www.thecellcollective.org | [25] |
| CellNetAnalyzer | Matlab® toolbox for exploring metabolic, signaling and regulatory networks. Uses Boolean and multivalued logical models and logic-based ODEs derived from Boolean models. Supports logical steady state analysis and computation of minimal intervention sets. | https://www2.mpi-magdeburg.mpg.de/projects/cna/cna.html | [26] |
| CellNOpt(R) / CytoCoptR | CellNetOptimizer (CNO): A Matlab® / R package for building logical models by training signaling networks derived from prior knowledge. | http://www.cellnopt.org https://github.com/cellnopt/CellNOptR http://www.cellnopt.org/cytocopter/ | [27,28] |
| CNORdt | CellNOpt add-on for training of a Boolean model from time-course data | https://bioconductor.org/packages/release/bioc/vignettes/CNORdt/inst/doc/CNORdt-vignette.pdf | --- |
| ChemChains | Cell Collective enables the construction, storage, as well as simulations of Boolean models. | http://www.bioinformatics.org/chemchains/wiki/ | [29] |
| CoLoMoTo | Diverse software tools developed by ColoMoto consortium members. | http://www.colomoto.org/software/ | [30] |

| CoLoMoTo Interactive Notebook | Workflow environment for the analysis of Boolean networks. | https://github.com/colomoto/colomoto-docker  https://colomoto.github.io/colomoto-docker/ | [31,32] |
|---|---|---|---|
| DDLab | Discrete Dynamics Laboratory, a tool for researching cellular automata and random Boolean networks. | http://www.ddlab.com | [33] |
| EpiLog | Boolean models for epithelial pattern formation | http://epilog-tool.org | [34] |
| FALCON | Toolbox for the Fast Contextualization of Logical Networks. | https://github.com/sysbiolux/FALCON | [35] |
| GDSCalc | Graph Dynamical Systems Calculator: A web-based application for Discrete Graph Dynamical Systems. | http://taos.vbi.vt.edu/gdscalc/ | [36] |
| GeneFAtt | C++ program for computing all attractors in synchronous and asynchronous Boolean networks. | https://sites.google.com/site/desheng619/download | [37] |
| GINSim | Gene Interaction Network simulation. Allows analyzing the qualitative dynamic behavior of GRN's based on a discrete, logical formalism. | http://ginsim.org | [38,39] |
| Jimena | Simulation framework for polynomial interpolation of Boolean networks using Boolean-tree-based data structures. | http://stefan-karl.de/jimena/ | [40] |
| JSBML | Java library for accessing SBML-qual models | http://sbml.org/Software/JSBML | [41] |
| kali | Calculates attrators of Boolean networks | https://github.com/arnaudporet/kali | [42,43] |
| Logical | Pipeline of computational methods for logical modelling of | https://github.com/sysbio-curie/Logical_modelling_pipeline | [44] |

| Modelling Pipeline | biological networks deregulated in diseases | | |
|---|---|---|---|
| MaBoSS 2 | Stochastic Boolean Markov modelling environment | https://maboss.curie.fr | [45] |
| NetBuilder | A graphical representation and simulation tool for logical models. | http://homepages.stca.herts.ac.uk/~erdqmjs/NetBuilder%20home/NetBuilder/ | [46] |
| NeuroDNet – Boolean Analysis | Converting a Boolean network into matrices for computation. | http://bioschool.iitd.ac.in/NeuroDNet/boolean.php | [47] |
| odefy | Transformation of Boolean models into ODEs. | https://www.helmholtz-muenchen.de/icb/software/odefy/index.html | [48] |
| optPBN | Matlab® optimization toolbox for probabilistic Boolean networks, an extension to the BN/PBN toolbox. | http://sourceforge.net/projects/optpbn/ | [49] |
| BN/PBN | Boolean and Probabilistic Boolean network Matlab® toolbox. | https://code.google.com/p/pbn-matlab-toolbox/downloads/list | --- |
| PATHLOGIC-S | Boolean framework for modelling of signaling networks. | https://sourceforge.net/projects/pathlogic/ | [50] |
| PDIC | Partial Information Decomposition and Context, an efficient network inference algorithm. | https://github.com/Tchanders/NetworkInference.jl | [51] |
| PHONEMeS (PHOsphorylation NEtworks | Modeling of Boolean signaling networks based on phosphoproteomics MS data | https://saezlab.github.io/PHONEMeS/ | [52] |

| for MS) | | | |
|---------|---------------------------------------------------------------------|----------------------------------------------------------------------------------|----------|
| Pint | Analysis tool for the dynamics of Boolean and multi-valued networks | https://hal.archives-ouvertes.fr/hal-01589248/document | [53] |
| Polynome | Construction of Boolean network models based on experimental time course data. | http://polymath.vbi.vt.edu/polynome/  https://github.com/jsjohnst/polynomevt | [54] |
| PolyBoRi | POLYnomials over BOolean Rings. | http://polybori.sourceforge.net | [55] |
| PROFILE | Personalization of patient-specific logical models | https://github.com/sysbio-curie/PROFILE | [56] |
| ProMoT | PROcess Modeling Tool for the construction of logical biochemical networks. | http://www2.mpi-magdeburg.mpg.de/projects/promot | [57] |
| PRUNET | Iterative network pruning based of a prior knowledge network (PKN) and Boolean expression profiles of stable phenotypes to deliver pruned contextualied networks. | http://prunet.sourceforge.net | [58] |
| pybool | Python package for inferring Boolean networks given a set of constraints. | https://pypi.python.org/pypi/pybool  https://github.com/JohnReid/pybool | [59] |
| PyBoolNet | Python package for the generation, modification and analysis of Boolean networks. Formerly known as BoolNetFixPoints. Uses BoolNet to compute all maximal symbolic steady states. | http://sourceforge.net/projects/boolnetfixpoints/ | [60,61] |
| RBN | Random Boolean network toolbox for Matlab® for simulation and visualization of RBNs. | http://www.mathworks.com/matlabcentral/fileexchange/3231-random-boolean-network-toolbox | --- |

| | | http://fias.uni-frankfurt.de/~willadsen/RBN http://www.teuscher.ch/rbntoolbox/ | |
|---|---|---|---|
| RBNLab | Software for studying the properties of different types of random Boolean networks. | http://rbn.sourceforge.net http://turing.iimas.unam.mx/~cgg/rbn/ | [62] |
| REACT | Reverse Engineering with Evolutionary Computational Tools. | https://github.com/veralicona/REACT | [63] |
| rxncon | A framework for visualizing and modelling of cellular networks derived from experimental data. | www.rxncon.org | [64,65] |
| SCNS | Single Cell Network Synthesis toolkit for the synthesis of Boolean GRNs. | http://scns.stemcells.cam.ac.uk | [66] |
| SELDOM | enSEmbLe of Dynamic lOgic-based Models, a R package for reverse engineering of signaling pathways | http://doi.org/10.5281/zenodo.250558 | [67] |
| SimBoolNet | Cytoscape plugin for Boolean simulation of signal transduction dynamics. | http://apps.cytoscape.org/apps/simboolnet | [68] |
| SMBioNet | Modelling of GRNs with multivalued logical functions combined with temporal logic. | http://www.i3s.unice.fr/~richard/smbionet/ | [140] |
| SQUAD | Derives continuous dynamical models from logical models and allows their simulation. | http://www.colomoto.org/software/squad.html | [69] |
| StableMotifs | Java library for attractor finding and control of Boolean networks. | https://github.com/jgtz/StableMotifs/ | [12,70] |
| STP | Matlab® toolbox for the Semi-Tensor Product of matrices, which | http://lsc.amss.ac.cn/~dcheng/stp/ | --- |

| | can be used to model Boolean networks [98]. | | |
|---|---|---|---|
| UpdateLabel | Algorithm enumerating all deterministic update schemes in Boolean Networks. | http://www.inf.udec.cl/~lilian/UDE/ | [52,71] |
| VisiBool | A tool for modelling, simulation and visualization of Boolean networks. | http://sysbio.uni-ulm.de/?Software:ViSiBooL | [72] |
| XBOOLE | Library for Boolean matrices. Useful for implementing Zhegalkin polynomial based models. | http://www.informatik.tu-freiberg.de/xboole/index.php | [73] |

**References:**

1.  Hinkelmann F, Brandon M, Guang B *et al.* ADAM: Analysis of Discrete Models of Biological Systems Using Computer Algebra. *Bmc Bioinformatics*, 12 (2011).
2.  Mizera A, Pang J, Su C, Yuan Q. ASSA-PBN: A Toolbox for Probabilistic Boolean Networks. *IEEE/ACM Trans Comput Biol Bioinform*, 15(4), 1203-1216 (2018).
3.  Mushthofa M, Torres G, Van de Peer Y, Marchal K, De Cock M. ASP-G: an ASP-based method for finding attractors in genetic regulatory networks. *Bioinformatics*, 30(21), 3086-3092 (2014).
4.  Shah OS, Chaudhary MFA, Awan HA *et al.* ATLANTIS - Attractor Landscape Analysis Toolbox for Cell Fate Discovery and Reprogramming. *Sci Rep*, 8(1), 3554 (2018).
5.  Naldi A. BioLQM: A Java Toolkit for the Manipulation and Conversion of Logical Qualitative Models of Biological Networks. *Front Physiol*, 9, 1605 (2018).
6.  Najafi A, Bidkhori G, Bozorgmehr JH, Koch I, Masoudi-Nejad A. Genome scale modeling in systems biology: algorithms and resources. *Current genomics*, 15(2), 130-159 (2014).
7.  Gerasch A, Faber D, Küntzer J *et al.* BiNA: a visual analytics tool for biological network data. *PLoS One*, 9(2), e87397 (2014).
8.  Mischnik M, Boyanova D, Hubertus K *et al.* A Boolean view separates platelet activatory and inhibitory signalling as verified by phosphorylation monitoring including threshold behaviour and integrin modulation. *Molecular bioSystems*, 9(6), 1326-1339 (2013).
9.  Hu J, Rho HS, Newman RH *et al.* Global analysis of phosphorylation networks in humans. *Biochimica et biophysica acta*, 1844(1 Pt B), 224-231 (2014).
10. Pirkl M, Hand E, Kube D, Spang R. Analyzing synergistic and non-synergistic interactions in signalling pathways using Boolean Nested Effect Models. *Bioinformatics*, 32(3), 409-416 (2015).
11. Dubrova E, Teslenko M. A SAT-based algorithm for finding attractors in synchronous Boolean networks. *IEEE/ACM transactions on computational biology and bioinformatics / IEEE, ACM*, 8(5), 1393-1399 (2011).
12. Zanudo JG, Albert R. Cell fate reprogramming by control of intracellular network dynamics. *PLoS Comput Biol*, 11(4), e1004193 (2015).
13. Albert I, Thakar J, Li S, Zhang R, Albert R. Boolean network simulations for life scientists. *Source code for biology and medicine*, 3, 16 (2008).
14. Benso A, Di Carlo S, Politano G, Savino A, Vasciaveo A. An extended gene protein/products Boolean network model including post-transcriptional regulation. *Theoretical biology & medical modelling*, 11 Suppl 1, S5 (2014).
15. Bock M, Scharp T, Talnikar C, Klipp E. BooleSim: an interactive Boolean network simulator. *Bioinformatics*, 30(1), 131-132 (2014).
16. Mcclenny LD, Imani M, Braga-Neto UM. BoolFilter: an R package for estimation and identification of partially-observed Boolean dynamical systems. *BMC Bioinformatics*, 18(1), 519 (2017).
17. Mussel C, Hopfensitz M, Kestler HA. BoolNet-an R package for generation, reconstruction and analysis of Boolean networks. *Bioinformatics*, 26(10), 1378-1380 (2010).
18. Lim CY, Wang H, Woodhouse S *et al.* BTR: training asynchronous Boolean models using single-cell expression data. *BMC Bioinformatics*, 17(1), 355 (2016).

19. Anderson J, Chang YC, Papachristodoulou A. Model decomposition and reduction tools for large-scale networks in systems biology. *Automatica*, 47(6), 1165-1174 (2011).

20. Albert R, Thakar J. Boolean modeling: a logic-based dynamic approach for understanding signaling and regulatory networks and for making useful predictions. *Wiley interdisciplinary reviews. Systems biology and medicine*, 6(5), 353-369 (2014).

21. Paroni A, Graudenzi A, Caravagna G, Damiani C, Mauri G, Antoniotti M. CABeRNET: a Cytoscape app for augmented Boolean models of gene regulatory NETworks. *BMC Bioinformatics*, 17(1), 64 (2016).

22. Correia RB, Gates AJ, Wang X, Rocha LM. CANA: A Python Package for Quantifying Control and Canalization in Boolean Networks. *Front Physiol*, 9, 1046 (2018).

23. Videla S, Saez-Rodriguez J, Guziolowski C, Siegel A. caspo: a toolbox for automated reasoning on the response of logical signaling networks families. *Bioinformatics*, 33(6), 947-950 (2017).

24. Guziolowski C, Videla S, Eduati F *et al.* Exhaustively characterizing feasible logic models of a signaling network using Answer Set Programming. *Bioinformatics*, 29(18), 2320-2326 (2013).

25. Helikar T, Kowal B, McClenathan S *et al.* The Cell Collective: toward an open and collaborative approach to systems biology. *BMC systems biology*, 6, 96 (2012).

26. Samaga R, Klamt S. Modeling approaches for qualitative and semi-quantitative analysis of cellular signaling networks. *Cell communication and signaling : CCS*, 11(1), 43 (2013).

27. Terfve C, Cokelaer T, Henriques D *et al.* CellNOptR: a flexible toolkit to train protein signaling networks to data using multiple logic formalisms. *BMC systems biology*, 6 (2012).

28. Saez-Rodriguez J, Alexopoulos LG, Epperlein J *et al.* Discrete logic modelling as a means to link protein signalling networks with functional analysis of mammalian signal transduction. *Mol Syst Biol*, 5, 331 (2009).

29. Helikar T, Rogers JA. ChemChains: a platform for simulation and analysis of biochemical networks aimed to laboratory scientists. *BMC systems biology*, 3, 58 (2009).

30. Naldi A, Monteiro PT, Mussel C *et al.* Cooperative development of logical modelling standards and tools with CoLoMoTo. *Bioinformatics*, 31(7), 1154-1159 (2015).

31. Naldi A, Hernandez C, Levy N *et al.* The CoLoMoTo Interactive Notebook: Accessible and Reproducible Computational Analyses for Qualitative Biological Networks. *Front Physiol*, 9, 680 (2018).

32. Levy N, Naldi A, Hernandez C *et al.* Prediction of Mutations to Control Pathways Enabling Tumor Cell Invasion with the CoLoMoTo Interactive Notebook (Tutorial). *Front Physiol*, 9, 787 (2018).

33. Wuensche A. *Exploring discrete dynamics* (Luniver Press, 2011).

34. Varela PL, Ramos CV, Monteiro PT, Chaouiya C. EpiLog: A software for the logical modelling of epithelial dynamics. *F1000Research*, (7), 9 (2018).

35. De Landtsheer S, Trairatphisan P, Lucarelli P, Sauter T. FALCON: a toolbox for the fast contextualization of logical networks. *Bioinformatics*, 33(21), 3431-3436 (2017).

36. Elmeligy Abdelhamid SH, Kuhlman CJ, Marathe MV, Mortveit HS, Ravi SS. GDSCalc: A Web-Based Application for Evaluating Discrete Graph Dynamical Systems. *PloS one*, 10(8), e0133660 (2015).

37. Zheng D, Yang G, Li X, Wang Z, Liu F, He L. An efficient algorithm for computing attractors of synchronous and asynchronous Boolean networks. *PloS one*, 8(4), e60593 (2013).

38. Naldi A, Berenguier D, Faure A, Lopez F, Thieffry D, Chaouiya C. Logical modelling of regulatory networks with GINsim 2.3. *Bio Systems*, 97(2), 134-139 (2009).

39. Naldi A, Hernandez C, Abou-Jaoudé W, Monteiro PT, Chaouiya C, Thieffry D. Logical Modeling and Analysis of Cellular Regulatory Networks With GINsim 3.0. *Front Physiol*, 9, 646 (2018).

40. Karl S, Dandekar T. Jimena: efficient computing and system state identification for genetic regulatory networks. *BMC bioinformatics*, 14, 306 (2013).

41. Rodriguez N, Thomas A, Watanabe L *et al.* JSBML 1.0: providing a smorgasbord of options to encode systems biology models. *Bioinformatics*, 31(20), 3383-3386 (2015).

42. Poret A, Boissel JP. An in silico target identification using Boolean network attractors: Avoiding pathological phenotypes. *C R Biol*, 337(12), 661-678 (2014).

43. Poret A, Guziolowski C. Therapeutic target discovery using Boolean network attractors: improvements of kali. *R Soc Open Sci*, 5(2), 171852 (2018).

44. Montagud A, Traynard P, Martignetti L *et al.* Conceptual and computational framework for logical modelling of biological networks deregulated in diseases. *Brief Bioinform*,    (2017).

45. Stoll G, Caron B, Viara E *et al.* MaBoSS 2.0: an environment for stochastic Boolean modeling. *Bioinformatics*, 33(14), 2226-2228 (2017).

46. Brown CT, Rust AG, Clarke PJ *et al.* New computational approaches for analysis of cis-regulatory networks. *Developmental biology*, 246(1), 86-102 (2002).

47. Vasaikar SV, Padhi AK, Jayaram B, Gomes J. NeuroDNet - an open source platform for constructing and analyzing neurodegenerative disease networks. *BMC Neurosci*, 14, 3 (2013).

48. Krumsiek J, Pölsterl S, Wittmann DM, Theis FJ. Odefy--from discrete to continuous models. *BMC Bioinformatics*, 11, 233 (2010).

49. Trairatphisan P, Mizera A, Pang J, Tantar AA, Sauter T. optPBN: an optimisation toolbox for probabilistic Boolean networks. *PloS one*, 9(7), e98001 (2014).

50. Fearnley LG, Nielsen LK. PATHLOGIC-S: a scalable Boolean framework for modelling cellular signalling. *PLoS One*, 7(8), e41977 (2012).

51. Chan TE, Stumpf MPH, Babtie AC. Gene Regulatory Network Inference from Single-Cell Data Using Multivariate Information Measures. *Cell Syst*, 5(3), 251-267.e253 (2017).

52. Terfve CD, Wilkes EH, Casado P, Cutillas PR, Saez-Rodriguez J. Large-scale models of signal propagation in human cells derived from discovery phosphoproteomic data. *Nat Commun*, 6, 8033 (2015).

53. Paulevé L. Pint: A Static Analyzer for Transient Dynamics of Qualitative Networks with IPython Interface. In: *CMSB 2017 - 15th conference on Computational Methods for Systems Biology.* Feret, J, Koeppl, H (Ed.^(Eds) (**Springer International Publishing**, Darmstadt, 2017) 370-376.

54. Dimitrova E, Garcia-Puente LD, Hinkelmann F *et al.* Parameter estimation for Boolean models of biological networks. *Theoretical Computer Science*, 412(26), 2816-2826 (2011).

55. Brickenstein M, Dreyer A. POLYBORI: A framework for Grobner-basis computations with Boolean polynomials. *J Symb Comput*, 44(9), 1326-1345 (2009).

56. Béal J, Montagud A, Traynard P, Barillot E, Calzone L. Personalization of Logical Models With Multi-Omics Data Allows Clinical Stratification of Patients. *Front Physiol*, 9, 1965 (2018).

57. Saez-Rodriguez J, Mirschel S, Hemenway R, Klamt S, Gilles ED, Ginkel M. Visual setup of logical models of signaling and regulatory networks with ProMoT. *BMC Bioinformatics*, 7, 506 (2006).

58. Rodriguez A, Crespo I, Androsova G, del Sol A. Discrete Logic Modelling Optimization to Contextualize Prior Knowledge Networks Using PRUNET. *PLoS One*, 10(6), e0127216 (2015).

59. Reid J. pybool: A Python package to infer Boolean networks under constraints.    (2011).

60. Klarner H, Bockmayr A, Siebert H. Computing Symbolic Steady States of Boolean Networks. *Cellular Automata: 11th International Conference on Cellular Automata for Research and Industry*, 8751, 561-570 (2014).

61. Klarner H, Streck A, Siebert H. PyBoolNet: a python package for the generation, analysis and visualization of boolean networks. *Bioinformatics*, 33(5), 770-772 (2017).

62. Gershenson C. Guiding the self-organization of random Boolean networks. *Theory Biosci*, 131(3), 181-191 (2012).

63. Vera-Licona P, Jarrah A, Garcia-Puente LD, McGee J, Laubenbacher R. An algebra-based method for inferring gene regulatory networks. *BMC systems biology*, 8, 37 (2014).

64. Mori T, Flöttmann M, Krantz M, Akutsu T, Klipp E. Stochastic simulation of Boolean rxncon models: towards quantitative analysis of large signaling networks. *BMC Syst Biol*, 9, 45 (2015).

65. Rother M, Münzner U, Thieme S, Krantz M. Information content and scalability in signal transduction network reconstruction formats. *Mol Biosyst*, 9(8), 1993-2004 (2013).

66. Moignard V, Woodhouse S, Haghverdi L *et al.* Decoding the regulatory network of early blood development from single-cell gene expression measurements. *Nat Biotechnol*, 33(3), 269-276 (2015).

67. Henriques D, Villaverde AF, Rocha M, Saez-Rodriguez J, Banga JR. Data-driven reverse engineering of signaling pathways using ensembles of dynamic models. *PLoS Comput Biol*, 13(2), e1005379 (2017).

68. Zheng J, Zhang D, Przytycki PF, Zielinski R, Capala J, Przytycka TM. SimBoolNet--a Cytoscape plugin for dynamic simulation of signaling networks. *Bioinformatics*, 26(1), 141-142 (2010).

69. Di Cara A, Garg A, De Micheli G, Xenarios I, Mendoza L. Dynamic simulation of regulatory networks using SQUAD. *BMC bioinformatics*, 8 (2007).

70. Pirkl M, Hand E, Kube D, Spang R. Analyzing synergistic and non-synergistic interactions in signalling pathways using Boolean Nested Effect Models. *Bioinformatics*,    (2015).

71. Palma E, Salinas L, Aracena J. Enumeration and extension of non-equivalent deterministic update schedules in Boolean networks. *Bioinformatics*, 32(5), 722-729 (2016).

72. Schwab J, Burkovski A, Siegle L, Müssel C, Kestler HA. ViSiBooL-visualization and simulation of Boolean networks with temporal constraints. *Bioinformatics*, 33(4), 601-604 (2017).

73. Posthoff CS, B. *Logic functions and equations: Binary models for computer science* (Springer, 2011).